\documentclass[11pt]{article}
\usepackage{amsmath}
\usepackage{amsfonts}
\usepackage{amssymb}
\begin{document}

\date{}
\title{\textbf{Symplectic geometries on supermanifolds}}
\author{\textsc{P.M.~Lavrov}\thanks{E-mail: lavrov@tspu.edu.ru} and
\textsc{O.V.~Radchenko}\thanks{E-mail: radchenko@tspu.edu.ru}\\
\\\textit{Tomsk State Pedagogical University,}
\\\textit{634041 Tomsk, Russia}}
\maketitle

\begin{quotation}
Extension of symplectic geometry on manifolds to the supersymmetric
case is considered. In the even case it leads to the even symplectic
geometry (or, equivalently, to the geometry on supermanifolds
endowed with a non-degenerate Poisson bracket) or to the geometry
on an even Fedosov supermanifolds. It is proven that in the odd case
there are two different scalar symplectic structures (namely, an odd
closed differential 2-form and the antibracket) which can be used
for construction of symplectic geometries on
supermanifolds. \noindent \normalsize
\end{quotation}

\vspace{.5cm}

\section{Introduction}

It is well-known that methods of symplectic geometry play very
important role in the formulation of classical mechanics on
manifolds (see, for example, \cite{Ar}). Deformation quantization
\cite{F} is formulated in terms of symplectic manifolds with a
symmetric connection compatible with the given symplectic structure
(the so-called Fedosov manifolds \cite{fm}). Formulation of
supersymmetric field theories, quantization of general gauge
theories introduced a number of applications of differential
geometry based on the notation of a supermanifold introduced and
studied by Berezin \cite{Ber}. In these cases, a supermanifold must
be endowed with an appropriate symplectic structure or (and) a
symmetric connection. Thus, investigating the geometrical contents
of the well-known Batalin-Vilkovisky quantization \cite{bv} is based
on using of the so-called antisymplectic supermanifolds which are
supermanifolds equipped with the antibracket \cite{geom}. In several
specific investigations in modern gauge field theory  \cite{btgl},
flat even Fedosov supermanifolds (in the terminology adopted here)
have been used.

Our aim of this work is to study extension of symplectic geometry on
manifolds to supersymmetric case. In the even case it leads to the
even symplectic geometry which is formulated on a supermanifold
equipped with an even closed non-degenerate differential 2-form (a
symplectic structure). It is equivalent to the geometry based on a
supermanifold equipped with the non-degenerate Poisson bracket. If,
in addition, a given symplectic supermanifold is endowed with a
symmetric connection (covariant derivative) compatible with the
symplectic structure, one has an even Fedosov supermanifolds which
can be considered as generalization of Fedosov manifolds \cite{fm}.
As for Fedosov manifolds the scalar curvature tensor for even
Fedosov supermanifolds vanishes. It is proven that in the odd case
there are two  scalar symplectic structures which can be
used for construction of symplectic geometries on
supermanifolds. First, one can equip a supermanifold with an odd
non-degenerate closed differential 2-forms to get an odd symplectic
supermanifold. Second, one can equip a supermanifold with the
antibracket (the antisymplectic structure) to get an antisymplectic
supermanifolds. Moreover, if we equip an odd symplectic supermanifold with a
symmetric connection compatible with the odd symplectic structure,
we get the geometry in which the scalar curvature tensor is
identically equal to zero. Situation is more interesting when an
antisymplectic supermanifold is equipped with a symmetric connection
compatible with a given antisymplectic structure. We prove that in
this case  the scalar curvature tensor is not, in
general, equal to zero. However,  the scalar curvature tensor squared is identically equal
to zero.

The paper is organised as follows. In Sect.~2, we  study
multiplication, contraction and symmetry properties  of tensor
fields on  supermanifolds. In Sect.~3, we consider scalar structures
which can be used for constructions of symplectic geometries on
supermanifolds. In Sect.~4, we discuss symmetric affine connections
and properties of their curvature tensors  on supermanifolds.  In
Sect.~5, we present the notion of the even symplectic geometry. In
Sect.~6, we introduce the notions of the odd symplectic geometry and
the antisymplectic geometry and study their basic properties.  
In Sect.~7, we give a short summary.

We use the condensed notation suggested by DeWitt \cite{DeWitt} and
definitions and notations adopted in \cite{lr}. Derivatives with
respect to the coordinates $x^i$ are understood as acting from the
left and are standardly denoted by ${\partial A}/{\partial
x^i}$. Right derivatives with respect to $x^i$ are labeled by the
subscript $"r"$ or the notation $A_{,i}={\partial_r A}/{\partial
x^i}$ is used. The Grassmann parity of any quantity $A$ is denoted
by $\epsilon (A)$.
\\

\section{Tensor fields}

In this section, we give  the basic definitions and relations in
tensor calculus on supermanifolds, to be  used in calculations in
what follows.

Let the variables $x^i, \epsilon(x^i)=\epsilon_i$ be local
coordinates on a supermanifold $M, dim M=N,$ in the vicinity of a
point $P\in M$. Let the sets $\{e_i\}$ and  $\{e^i\}$ be coordinate
bases in the tangent space $T_PM$ and the cotangent space $T^*_PM$,
respectively. If one goes over to another set ${\bar x}^{i}={\bar
x}^i(x)$ of local coordinates, the basis vectors in $T_PM$ and
$T^*_PM$ transform as
\begin{eqnarray}
\label{vec}
 {\bar e}_i=e_j \frac{\partial_r x^j}{\partial {\bar x}^i},
 \quad
{\bar e}^i=e^j \frac{\partial {\bar x}^i}{\partial x^j}.
\end{eqnarray}
For the transformation matrices the following relations hold:
\begin{eqnarray}
\label{unitJ}
 \frac{\partial_r {\bar x}^i}{\partial x^k}
 \frac{\partial_r x^k}{\partial {\bar x}^j}=\delta^i_j,
 \quad
 \frac{\partial x^k}{\partial {\bar x}^j}
 \frac{\partial {\bar x}^i}{\partial x^k}=\delta^i_j,
 \quad
 \frac{\partial_r x^i}{\partial {\bar x}^k}
 \frac{\partial_r {\bar x}^k}{\partial  x^j}=\delta^i_j,
 \quad
 \frac{\partial {\bar x}^k}{\partial  x^j}
 \frac{\partial  x^i}{\partial {\bar x}^k}=\delta^i_j.
\end{eqnarray}

A tensor field of type $(n,m)$ and rank $n+m$ is defined as a
geometric object given by a set of functions with $n$ upper
and $m$ lower indices in each local coordinate system
$(x)=(x^1,...,x^N)$ with certain transformation laws. We omit
the general definition (see \cite{lr}) and restrict ourself to cases
of vector fields $T^i$ and co-vector fields $T_i$
\begin{eqnarray}
\label{formvec}
 {\bar T}^i= T^n\frac{\partial {\bar x}^i}{\partial x^n}\,, 
 \qquad
 {\bar T}_i= T_n\frac{\partial_r x^n}{\partial{\bar x}^i}
\end{eqnarray}
and of second-rank tensor fields of different types
\begin{eqnarray}
\label{formup}
{\bar T}^{ij}&=&
T^{mn}\frac{\partial {\bar x}^j}{\partial x^n}
\frac{\partial {\bar x}^i}{\partial x^m}
(-1)^{\epsilon_j(\epsilon_i+\epsilon_m)},\\
\label{form}
{\bar T}_{ij}&=&
T_{mn}\frac{\partial_r x^n}{\partial {\bar x}^j}
\frac{\partial_r x^m}{\partial {\bar x}^i}
(-1)^{\epsilon_j(\epsilon_i+\epsilon_m)},\\
\label{form1} {\bar T}^i_{\;\;j}&=& T^m_{\;\;\;n}\frac{\partial_r
x^n}{\partial {\bar x}^j} \frac{\partial {\bar x}^i}{\partial x^m}
(-1)^{\epsilon_j(\epsilon_i+\epsilon_m)}\,.
\end{eqnarray}
Note that the unit matrix $\delta^i_j$ is related to the unit tensor
field $E^i_{\;j}$  transforming in accordance with (\ref{form1}) as
\begin{eqnarray}
\label{unit} E^i_{\;j}=\delta^i_j.
\end{eqnarray}

From a tensor field of type $(n,m)$ and rank $n+m$, where $n\neq 0,
\;m\neq 0$, one can construct a tensor field of type $(n-1,m-1)$ and
rank $n+m-2$ by the contraction of an upper and a lower index by the
rules (for details, see \cite{lr}). In particular, for tensor
fields of type $(1,1)$, the contraction gives the supertrace,
\begin{eqnarray}
\label{sc1} T^i_{\;\;i}\;(-1)^{\epsilon_i}.
\end{eqnarray}
Using the multiplication operation, from two tensor fields of types
$(n,0)$ and $(0,m)$, one can construct new tensor fields of type
$(n-1,m-1)$. In particular, vector $U^i$ and covector $V_i$ fields
thus yield a scalar field
\begin{eqnarray}
\label{sc}
 (-1)^{\epsilon_i(\epsilon({\bf V})+1)}\;U^i\;V_i =
 (-1)^{\epsilon({\bf U})\epsilon({\bf V}) +
 \epsilon_i\epsilon({\bf U})}\;V_i\;U^i\,,
\end{eqnarray}
which is invariant with respect to the choice of local coordinates.
Two second-rank tensor fields $U^{ij}$ and $V_{ij}$ yield the tensor
fields
\begin{eqnarray}
\label{contr} (-1)^{(\epsilon_i+\epsilon_k)\epsilon({\bf
V})+\epsilon_k}\;U^{ik}\;V_{kj} \quad {\rm and} \quad
(-1)^{(\epsilon_i+\epsilon_k)\epsilon({\bf
V})+\epsilon_k(\epsilon_i+\epsilon_j+1)}\;U^{ki}V_{jk}
\end{eqnarray}
transforming in accordance with  (\ref{form1}).  Further contracting
indices yields the scalar
\begin{eqnarray}
\label{contr3} (-1)^{(\epsilon_i+\epsilon_k)(\epsilon({\bf
V})+1)}\;U^{ik}\;V_{ki} = (-1)^{\epsilon({\bf U})\epsilon({\bf V})+
(\epsilon_i+\epsilon_k)\epsilon({\bf U})}\;V_{ik}\;U^{ki}\,.
\end{eqnarray}
Moreover, recalling (\ref{unit}) and (\ref{contr}), the  inverse
tensor field $T_{ij}$  for a non-degenerate  second-rank tensor field
$T^{ij}$ of type $(2,0)$ should be defined via the relations
\begin{eqnarray}
\label{invers1} &&(-1)^{(\epsilon_i+\epsilon_k)\epsilon({\bf
T})+\epsilon_k}\;T^{ik}\;T_{kj}
= \delta^i_j\,,\\
\label{invers2} &&(-1)^{(\epsilon_j+\epsilon_k)\epsilon({\bf
T})+\epsilon_j}\;T_{jk}\;T^{ki} = \delta^i_j\,,
\\
\nonumber
 &&\epsilon(T_{ij})=\epsilon(T^{ij})=
 \epsilon({\bf T})+\epsilon_i+\epsilon_j\,,
\end{eqnarray}
and similarly for tensor fields of type (0,2).

It is well known that in constructing a tensor calculus on manifold,
an important role is played by symmetric and antisymmetric tensor
fields. In the supersymmetric case, supermatrices have more possible
symmetry properties (eight types \cite{GT}), and a natural question
is whether these properties are compatible with the tensor
transformation laws. Among the eight types of supermatrices with
possible symmetry properties there exist only two ones satisfying
tensor transformation laws. In our definition of tensor fields on
supermanifolds, only the supermatrices having the generalized
symmetry or antisymmetry properties satisfy the tensor
transformation laws. Indeed, let us consider a second-rank
supermatrix of type $(2,0)$ having the generalized symmetry
(antisymmetry) property
\begin{eqnarray}
\label{sym}
 T^{ij}=(-1)^{\epsilon_i\epsilon_j}T^{ji}\quad
 (T^{ij}=-(-1)^{\epsilon_i\epsilon_j}T^{ji}).
\end{eqnarray}
 This property is compatible with the transformation law
(\ref{formup}),
\begin{eqnarray}
\nonumber {\bar T}^{ij}= T^{mn}\frac{\partial {\bar x}^j}{\partial
x^n} \frac{\partial {\bar x}^i}{\partial x^m}
(-1)^{\epsilon_j(\epsilon_i+\epsilon_m)}= T^{nm} \frac{\partial
{\bar x}^i}{\partial x^m} \frac{\partial {\bar x}^j}{\partial x^n}
(-1)^{\epsilon_i\epsilon_n} =(-1)^{\epsilon_i\epsilon_j}{\bar
T}^{ji}
\end{eqnarray}
and similarly for antisymmetry property. Other possible symmetry
types of supermatrices do not survive verification of the
compatibility with adopted tensor transformation laws. We note that
for non-degenerate symmetric and antisymmetric tensor fields, their
inverse tensor fields also have the necessary symmetry properties.
For example, we consider a second-rank tensor field $T^{ij}$ with
symmetry (antisymmetry) property. From definition (\ref{invers1})
and (\ref{invers2}), we can then find that the inverse tensor field
\begin{eqnarray}
\label{syminv}
 T_{ij}=(-1)^{\epsilon_i\epsilon_j+\epsilon(T)}T_{ji}\quad
 (T_{ij}=-(-1)^{\epsilon_i\epsilon_j+\epsilon(T)}T_{ji})
\end{eqnarray}
also has the generalized symmetry (antisymmetry) property.

\section{Scalar structures on supermanifolds}
In this section, we discuss  important scalar structures on
supermanifolds which can be defined in terms of symmetric and
antisymmetric tensor fields. Namely, we are going to consider
definitions and basic properties of the super-Poisson bracket, the
antibracket and the differential 2-form which are main objects in
formulation of Quantum Field Theory.

The Poisson bracket is defined on a supermanifold $M$ with even
dimension for any two scalar functions $A$ and $B$ as an even
bilinear operation $\{A,B\}$, $\epsilon(\{A,B\})=\epsilon({\bf
A})+\epsilon({\bf
 B})$ having the generalized antisymmetry property
\begin{eqnarray}
\label{SPBsym} \{A,B\}=-(-1)^{\epsilon({\bf A})\epsilon({\bf
B})}\{B,A\}
\end{eqnarray}
and obeying the Jacobi identity
\begin{eqnarray}
\label{JIPBsym} \{A,\{B,C\}\}(-1)^{\epsilon({\bf A})\epsilon({\bf
C})}+cycle(A,B,C)\equiv 0.
\end{eqnarray}
We can define the  Poisson bracket by the relation
\begin{eqnarray}
\label{PB}
\{A,B\}=\frac{\partial_r A}{\partial x^i}
\omega^{ij}\frac{\partial B}{\partial x^j},\quad \epsilon(\omega^{ij})
=\epsilon_i+\epsilon_j.
\end{eqnarray}
If $\omega^{ij}$ is a second-rank tensor field of type $(2,0)$ then
this definition gives the invariance of the Poisson bracket under
local coordinate transformations $x\rightarrow {\bar x},\; \{{\bar
A},{\bar B}\}=\{A,B\}$. If $\omega^{ij}$ has the generalized
antisymmetry property
\begin{eqnarray}
\label{omegaanti}
\omega^{ij}=-(-1)^{\epsilon_i\epsilon_j}\omega^{ji}
\end{eqnarray}
then the definition (\ref{PB}) reproduce the property (\ref{SPBsym}).
In terms of $\omega^{ij}$ the Jacobi identity means fulfilment of the
following relations
\begin{eqnarray}
\label{omegaJI} \omega^{in}\frac{\partial\omega^{jk}}{\partial
x^n}(-1)^{\epsilon_i\epsilon_k} +cycle(i,j,k)\equiv 0.
\end{eqnarray}
Now, suppose that the tensor field $\omega^{ij}$ is non-degenerate.
We can introduce the inverse tensor field $\omega_{ij}$ which has
also the generalized antisymmetry property
\begin{eqnarray}
\label{omegainvanti}
\omega_{ij}=-(-1)^{\epsilon_i\epsilon_j}\omega_{ji}.
\end{eqnarray}
In terms of $\omega_{ij}$, the Jacobi identity (\ref{omegaJI}) can
be rewritten in the form
\begin{eqnarray}
\label{omegainvJI}
\omega_{ij,k}(-1)^{\epsilon_i\epsilon_k}+cycle(i,j,k)\equiv 0.
\end{eqnarray}
The tensor field $\omega_{ij}$ defines the differential 2-form on
the supermanifold $M$
\begin{eqnarray}
\label{omegaform} \omega=\omega_{ij}\;dx^j\wedge dx^i,\quad
dx^j\wedge dx^i=-(-1)^{\epsilon_i\epsilon_j}dx^i\wedge dx^j, \quad
\epsilon(\omega)=0
\end{eqnarray}
which is invariant under a change of the local coordinates, ${\bar
\omega}=\omega$. The external derivative is given by
\begin{eqnarray}
\label{omegaformd} d\omega=\omega_{ij,k}\;dx^k\wedge dx^j\wedge
dx^i.
\end{eqnarray}
It is also invariant under a change of the local coordinates,
$d{\bar \omega}=d\omega$. Moreover, due to the identities
(\ref{omegainvJI}) the differential non-degenerate 2-form $\omega$
(\ref{omegaform}) is closed
\begin{eqnarray}
\label{omegaformcl} d\omega=0.
\end{eqnarray}
Therefore, any non-degenerate super-Poisson bracket on a
supermanifold defines an even non-degenerate closed differential
2-form and via verse.

The antibracket is defined for any two scalar functions $A$ and $B$
as an odd bilinear operation $(A,B)$, $\epsilon((A,B))=\epsilon({\bf
A})+\epsilon({\bf
 B})+1$ having the generalized antisymmetry property
\begin{eqnarray}
\label{ABsym} (A,B)=-(-1)^{(\epsilon({\bf A})+1)(\epsilon({\bf
B})+1)}(B,A)
\end{eqnarray}
and obeying the Jacobi identity
\begin{eqnarray}
\label{ABJIsym} (A,(B,C))(-1)^{(\epsilon({\bf A})+1)(\epsilon({\bf
C})+1)}+cycle(A,B,C)\equiv 0.
\end{eqnarray}
We can define the antibracket by the relation
\begin{eqnarray}
\label{AB} (A,B)=\frac{\partial_r A}{\partial x^i}(-1)^{\epsilon_i}
\Omega^{ij}\frac{\partial B}{\partial x^j},\quad
\epsilon(\Omega^{ij}) =\epsilon_i+\epsilon_j+1.
\end{eqnarray}
If $\Omega^{ij}$ is a second-rank tensor field of type $(2,0)$ then
this definition leads to the invariance of the  antibracket under
local coordinate transformations $x\rightarrow {\bar x},\; ({\bar
A},{\bar B})=(A,B)$. If $\Omega^{ij}$ has the generalized symmetry
property
\begin{eqnarray}
\label{Omegasym} \Omega^{ij}=(-1)^{\epsilon_i\epsilon_j}\Omega^{ji}
\end{eqnarray}
then the definition (\ref{AB}) reproduce the property (\ref{ABsym}).
In terms of $\Omega^{ij}$ the Jacobi identity means fulfilment of
the following relations
\begin{eqnarray}
\label{OmegaJI} \Omega^{in}\frac{\partial\Omega^{jk}}{\partial
x^n}(-1)^{\epsilon_i(\epsilon_k+1)} +cycle(i,j,k)\equiv 0.
\end{eqnarray}
When the tensor field $\Omega^{ij}$ (\ref{Omegasym}) is non-degenerate
then the inverse
tensor field $\Omega_{ij}$ has  the generalized antisymmetry property
\begin{eqnarray}
\label{Omegainvsym} \Omega_{ij}=-(-1)^{\epsilon_i\epsilon_j}\Omega_{ji},\quad
\epsilon(\Omega_{ij})= \epsilon_i+\epsilon_j+1.
\end{eqnarray}
In terms of $\Omega_{ij}$ the Jacobi identity (\ref{OmegaJI}) can be rewritten
in the form
\begin{eqnarray}
\label{OmegainvJI} \Omega_{ij,k}(-1)^{\epsilon_k(\epsilon_i+1)}
+cycle(i,j,k)\equiv 0.
\end{eqnarray}

We can also introduce an odd closed non-degenerate differential 2-form
on a supermanifold
by the relation
\begin{eqnarray}
\label{omegaoddform} \omega=\omega_{ij}\;dx^j\wedge dx^i,\quad
dx^j\wedge dx^i=-(-1)^{\epsilon_i\epsilon_j}dx^i\wedge dx^j, \quad
\epsilon(\omega)=1,\quad \omega_{ij}=-(-1)^{\epsilon_i\epsilon_j}\omega_{ji}
\end{eqnarray}
which has formally the same properties as in even case. In
particular, in terms of tensor field $\omega_{ij}$ the closure of
$\omega$ ($d\omega=0$) has the form (\ref{omegainvJI}). We see that
in the odd case any antibracket defines an odd closed differential
2-form and therefore an antisymplectic supermanifold should be considered as an odd symplectic supermanifold. 
\\

\section{Covariant derivatives and curvature tensor}

As in the case of tensor analysis on manifolds, on a supermanifold
$M$ one can introduce the covariant derivation (or affine
connection) as a mapping $\nabla$ (with components $\nabla_i,\,
\epsilon(\nabla_i)= \epsilon_i$) from the set of tensor fields on
$M$ to itself by the requirement that it should be a tensor
operation acting from the right and adding one more lower index and,
when it is possible locally to introduce Cartesian coordinates on
$M$, that it should reduce to the usual (right--)differentiation. It
allows to construct the action of covariant derivatives on tensor
fields of different types. In particular, they are given as local
operations acting on scalar, vector and co-vector fields
by the rules
\begin{eqnarray}
\label{scal} T\,\nabla_i&=&T_{,i}\,,
\\
\label{vector} T^i\,\nabla_j&=&T^i_{\;,j}+ T^k\Gamma^i_{\;kj}
(-1)^{\epsilon_k(\epsilon_i+1)}\,,
\\
T_i\,\nabla_j&=&T_{i,j}-T_k\Gamma^k_{\;ij}\,,
\end{eqnarray}
and on second-rank tensor fields of type $(2,0), (0,2)$ and $(1,1)$
by the rules
\begin{eqnarray}
{T}^{ij}\,{\nabla}_k&=& {T}^{ij}_{\;\;,k} +
{T}^{il}\,\Gamma^j_{\;lk}(-1)^{\epsilon_l(\epsilon_j+1)}+
{T}^{lj}\,\Gamma^i_{\;lk}
(-1)^{\epsilon_i\epsilon_j+\epsilon_l(\epsilon_i+\epsilon_j+1)}\,,\\
{T}_{ij}\,{\nabla}_k&=& {T}_{ij,k} - {T}_{il}\,\Gamma^l_{\;jk}-
{T}_{lj}\,\Gamma^l_{\;ik}
(-1)^{\epsilon_i\epsilon_j+\epsilon_l\epsilon_j}\,,\\
{T}^i_{\;j}\,{\nabla}_k&=& { T}^i_{\;\;j,k} -
{T}^i_{\;l}\,\Gamma^l_{\;jk} + {T}^l_{\;j}\,\Gamma^i_{\;lk}
(-1)^{\epsilon_i\epsilon_j+\epsilon_l(\epsilon_i+\epsilon_j+1)}\,.
\end{eqnarray}
Here, $\Gamma^i_{\;jk}$ are  affine connection components.
Similarly, the action of the covariant derivative on a tensor field
of any rank and type is given in terms of their tensor components,
their ordinary derivatives and the connection components.

In general, the connection components $\Gamma^i_{\;jk}$ do not have
the property of (generalized) symmetry w.r.t. the lower indices. The
deviation from this symmetry is the torsion,
\begin{eqnarray}
T^i_{\;jk} := \Gamma^i_{\;jk} -
(-1)^{\epsilon_j\epsilon_k}\Gamma^i_{\;kj}\,,
\end{eqnarray}
which transforms as a tensor field.
 If a supermanifold $M$ is torsionless, i.e., if a connection
obey the relation
\begin{eqnarray}
\label{Crisp} \Gamma^i_{\;jk}=
(-1)^{\epsilon_j\epsilon_k}\Gamma^i_{\;kj},
\end{eqnarray}
then one says that a symmetric connection is defined on $M$. Here,
we consider only symmetric connections.

The curvature tensor  $R^i_{\;\;mjk}$ of a given symmetric
connection is defined in a coordinate basis by the action of the
commutator of covariant derivatives, $[\nabla_i,\nabla_j]=
\nabla_i\nabla_j-(-1)^{\epsilon_i\epsilon_j}\nabla_j\nabla_i$, on a
vector field $T^i$ as
\begin{eqnarray}
\label{Rie} T^i[\nabla_j,\nabla_k]=-(-1)^{\epsilon_m(\epsilon_i+1)}
T^mR^i_{\;\;mjk}.
\end{eqnarray}
The choice of factor in r.h.s (\ref{Rie}) is dictated by the
requirement for product of tensor fields of types $(1,0)$ and
$(1,3)$ to be a tensor field of type $(1,2)$. A straightforward
calculation yields
\begin{eqnarray}
\label{R} R^i_{\;\;mjk}=-\Gamma^i_{\;\;mj,k}+
\Gamma^i_{\;\;mk,j}(-1)^{\epsilon_j\epsilon_k}+
\Gamma^i_{\;\;jn}\Gamma^n_{\;\;mk}(-1)^{\epsilon_j\epsilon_m}-
\Gamma^i_{\;\;kn}\Gamma^n_{\;\;mj}
(-1)^{\epsilon_k(\epsilon_m+\epsilon_j)}.
\end{eqnarray}
The curvature tensor field possesses the following generalized
antisymmetry property,
\begin{eqnarray}
\label{Rsym}
R^i_{\;\;mjk}=-(-1)^{\epsilon_j\epsilon_k}R^i_{\;\;mkj}\,;
\end{eqnarray}
furthermore, it obeys the  Jacobi identity,
\begin{eqnarray}
\label{Rjac} (-1)^{\epsilon_m\epsilon_k}R^i_{\;\;mjk}
+(-1)^{\epsilon_j\epsilon_m}R^i_{\;\;jkm}
+(-1)^{\epsilon_k\epsilon_j}R^i_{\;\;kmj}\equiv 0\,.
\end{eqnarray}
Using the  Jacobi identity for the covariant derivatives,
\begin{eqnarray}
\label{}
[\nabla_i,[\nabla_j,\nabla_k]](-1)^{\epsilon_i\epsilon_k}+
[\nabla_k,[\nabla_i,\nabla_j]](-1)^{\epsilon_k\epsilon_j}+
[\nabla_j,[\nabla_k,\nabla_i]](-1)^{\epsilon_i\epsilon_j}\equiv
0\,,
\end{eqnarray}
one obtains the  Bianchi identity,
\begin{eqnarray}
\label{BI} (-1)^{\epsilon_i\epsilon_j}R^n_{\;\;mjk;i}
+(-1)^{\epsilon_i\epsilon_k}R^n_{\;\;mij;k}
+(-1)^{\epsilon_k\epsilon_j}R^n_{\;\;mki;j}\equiv 0\,,
\end{eqnarray}
with the notation $R^n_{\;\;mjk;i}:\,=R^n_{\;\;mjk}\nabla_i$.
\\

\section{Even symplectic geometry}

Suppose now that we are given a supermanifold $M$ of an even
dimension, ${\rm dim}\, M=2n$. Let $\omega$ be an even
non-degenerate differeitial 2-form (\ref{omegaform}) on $M$. Then,
the pair $(M,\omega)$ is called an even  almost symplectic
supermanifold; it is called an even  symplectic supermanifold if
$\omega$ is closed, $d \omega = 0$. The inverse tensor field
$\omega^{ij}$ defines the non-degenarate Poisson bracket.
Supermanifolds equipped with this structure are called
non-degenarate Poisson supermanifolds. From the above considerations
it follows that, as in the case of ordinary symplectic geometry on
manifolds, there exists  one-to-one correspondence between an even
symplectic supermanifold and the non-degenerate Poisson
supermanifold.

 Let $\nabla$ (or $\Gamma$) be a covariant derivative
(a symmetric connection) on $M$ which preserves the 2-form $\omega$,
$\omega\nabla=0$.  In a coordinate basis this requirement reads
\begin{eqnarray}
\label{covomiv} \omega_{ij,k}-\omega_{im}\Gamma^m_{\;\;jk}+
\omega_{jm}\Gamma^m_{\;\;ik}(-1)^{\epsilon_i\epsilon_j}=0.
\end{eqnarray}
If, in addition, $\Gamma$ is symmetric then we have an even
symplectic connection (or symplectic covariant derivative) on $M$.
Now, an even Fedosov supermanifold $(M,\omega,\Gamma)$ is
defined as an even  symplectic supermanifold with a given
even symplectic connection.

Let us introduce the curvature tensor of an even symplectic
connection,
\begin{eqnarray}
\label{Rs}
R_{ijkl}=\omega_{in}R^n_{\;\;jkl},\quad
\epsilon(R_{ijkl})=\epsilon_i+
\epsilon_j+\epsilon_k+\epsilon_l,
\end{eqnarray}
where $R^n_{\;\;jkl}$ is given by (\ref{R}). This leads to the
following representation,
\begin{eqnarray}
\label{Rse} R_{imjk}=-\omega_{in}\Gamma^n_{\;\;mj,k}+
\omega_{in}\Gamma^n_{\;\;mk,j}(-1)^{\epsilon_j\epsilon_k}+
\Gamma_{ijn}\Gamma^n_{\;\;mk}(-1)^{\epsilon_j\epsilon_m}-
\Gamma_{ikn}\Gamma^n_{\;\;mj}
(-1)^{\epsilon_k(\epsilon_m+\epsilon_j)}\,,
\end{eqnarray}
where we used the notation
\begin{eqnarray}
\label{G} \Gamma_{ijk}=\omega_{in}\Gamma^n_{\;\;jk},\quad
\epsilon(\Gamma_{ijk})=
\epsilon_i+\epsilon_j+\epsilon_k\,.
\end{eqnarray}
Using this, the relation (\ref{covomiv}) reads
\begin{eqnarray}
\label{covom}
\omega_{ij,k}=\Gamma_{ijk}-
\Gamma_{jik}(-1)^{\epsilon_i\epsilon_j}.
\end{eqnarray}
Furthermore, from Eq.~(\ref{R}) it is obvious that
\begin{eqnarray}
\label{Rans} R_{ijkl}=-(-1)^{\epsilon_k\epsilon_l}R_{ijlk},
\end{eqnarray}
and, using (\ref{R}) and (\ref{Rjac}), one deduces the
Jacobi identity for $R_{ijkl}$,
\begin{eqnarray}
\label{Rjac1} (-1)^{\epsilon_j\epsilon_l}R_{ijkl}
+(-1)^{\epsilon_l\epsilon_k}R_{iljk}
+(-1)^{\epsilon_k\epsilon_j}R_{iklj}=0\,.
\end{eqnarray}
In addition, the curvature tensor $R_{ijkl}$ is generalized
symmetric w.r.t. the first two indices,
\begin{eqnarray}
\label{Ras1}
R_{ijkl}=(-1)^{\epsilon_i\epsilon_j}R_{jikl}.
\end{eqnarray}
In order to prove this, let us consider the relations which follow
from (\ref{covomiv})
\begin{eqnarray}
\label{com}
\omega_{ij,kl}=\Gamma_{ijk,l}-
\Gamma_{jik,l}(-1)^{\epsilon_i\epsilon_j}.
\end{eqnarray}
Then, using the relations
\begin{eqnarray}
\label{G1} \Gamma_{ijk,l}=\omega_{in}\Gamma^n_{\;\;jk,l}
+\omega_{in,l}\Gamma^n_{\;\;jk}
(-1)^{(\epsilon_n+\epsilon_j+\epsilon_k)\epsilon_l}
\end{eqnarray}
and the definitions (\ref{Rse}) and (\ref{covom}), we get
\begin{eqnarray}
\label{com1}
\nonumber
0&=&\omega_{ij,kl}-(-1)^{\epsilon_k\epsilon_l}\omega_{ij,lk}\\
\nonumber
&=&\Gamma_{ijk,l}-
\Gamma_{jik,l}(-1)^{\epsilon_i\epsilon_j}
-\Gamma_{ijl,k}(-1)^{\epsilon_k\epsilon_l}+
\Gamma_{jil,k}(-1)^{\epsilon_i\epsilon_j+\epsilon_k\epsilon_l}\\
&=&-R_{ijkl}+(-1)^{\epsilon_i\epsilon_j}R_{jikl}.
\end{eqnarray}

For any even  symplectic connection there holds the identity
\begin{eqnarray}
\label{Rjac2}
R_{ijkl}
+(-1)^{\epsilon_l(\epsilon_i+\epsilon_k+\epsilon_j)}R_{lijk}
+(-1)^{(\epsilon_k+\epsilon_l)(\epsilon_i+\epsilon_j)}
R_{klij}+
(-1)^{\epsilon_i(\epsilon_j+\epsilon_l+\epsilon_k)}R_{jkli}=0.
\end{eqnarray}
This is proved by using the Jacobi identity (\ref{Rjac1}) together
with a cyclic change of the indices (see \cite{lr}). The identity
(\ref{Rjac2}) involves components of the curvature tensor with cyclic
permutation of all indices, but the sign factors depending on the
Grassmann parities of the indices do not follow from a cyclic permutation,
as is the case, for example, for the Jacobi identity, but are defined by
the permutation of the indices that takes a given set into the original one.
 In the case of ordinary manifolds, i.e., when all the variables
 $x^i$ are even
($\epsilon_i=0$), Eq. (\ref{Rjac2}) obtains the symmetric form
\cite{fm},
\begin{eqnarray}
\label{Rjac3}
R_{ijkl} + R_{lijk} +R_{klij} +R_{jkli}=0.
\end{eqnarray}

Having  the curvature tensor, $R_{ijkl}$, and the inverse tensor
field $\omega^{ij}$, with allowance made for the symmetry properties
of these tensors, (\ref{omegaanti}), (\ref{Rans}) and (\ref{Ras1}),
one can construct  the only  tensor field of type $(0,2)$,
\begin{eqnarray}
 \label{R2} K_{ij}= \omega^{kn}R_{nikj}
 (-1)^{\epsilon_i\epsilon_k+\epsilon_k+\epsilon_n}
 \;=\;R^k_{\;\;ikj}\;(-1)^{\epsilon_k(\epsilon_i+1)},\quad
 \epsilon(K_{ij})=\epsilon_i+\epsilon_j.
\end{eqnarray}
This tensor has the generalized symmetry property
\begin{eqnarray}
\label{Rl3} K_{ij}=(-1)^{\epsilon_i\epsilon_j}K_{ji}
\end{eqnarray}
and is called the Ricci tensor.

Now we can define the scalar curvature tensor $K$ by the formula
\begin{eqnarray}
\label{Rsc} K=\omega^{ji}K_{ij}(-1)^{\epsilon_i+\epsilon_j}.
\end{eqnarray}
From the symmetry properties of $K_{ij}$ and $\omega^{ij}$, it
follows that
\begin{eqnarray}
\label{Rsc1} K=0.
\end{eqnarray}
Therefore, as in the case of Fedosov manifolds \cite{fm}, for any
even symplectic connection the scalar curvature tensor necessarily
vanishes.

\section{Odd symplectic geometry}

Consider now possible constructions of geometry on supermanifolds in
odd supersymmetric extension of symplectic geometry on manifolds. We
know that in the odd case there exist two independent structures
constructed with the help of generalized symmetric (an antibracket)
and antisymmetric (a 2-form) second-rank tensor fields. 

Suppose that  a supermanifold $M$ of an even dimension (${\rm dim}\,
M=2n$) is equipped both with an odd closed non-degenerate
differential 2-form
\begin{eqnarray}
\label{omegaoddform} \omega=\omega_{ij}\;dx^j\wedge dx^i,\quad
\omega_{ij}=-(-1)^{\epsilon_i\epsilon_j}\omega_{ji}, \quad
\epsilon(\omega)=1\;,\quad d\omega = 0
\end{eqnarray}
and a symmetric connection (covariant derivative) compatible with a
 given symplectic structure $\omega$
\begin{eqnarray}
\label{covomoddiv} \omega_{ij,k}-\omega_{im}\Gamma^m_{\;\;jk}+
\omega_{jm}\Gamma^m_{\;\;ik}(-1)^{\epsilon_i\epsilon_j}=0.
\end{eqnarray}
Repeatting all calculations of previous section and taking into account that
\begin{eqnarray}
\label{Rspar}
\epsilon(R_{ijkl})=\epsilon_i+
\epsilon_j+\epsilon_k+\epsilon_l+1,\quad \epsilon(\Gamma_{ijk})=
\epsilon_i+\epsilon_j+\epsilon_k+1\;,
\end{eqnarray}
we obtain that all relations and identities for the curvature
tensor have the same
forms as in the case of even symplectic supermanifolds. There are two essential differences only. The first one is connected with Ricchi tensor which has no special symmetry properties. The second one is non-triviality of the scalar curvature tensor. Both these statements will be considered below within antisymplectic supermanifolds. 

Consider the second possibility to construct an odd symplectic
geometry. To do this let us equip a supermanifold $M$ with an
antibracket (\ref{AB}). In its turn, let tensor field $\Omega^{ij}$
( an antisymplectic structure) be covariant constant
\begin{equation}
\label{Omegacon}
\Omega^{ij}\nabla_k=0.
\end{equation}
Then the inverse tensor field $\Omega_{ij}$ will be covariant constant too
\begin{equation}
\label{Omegainvcon} \Omega_{ij}\nabla_k=0\;, \quad
\Omega_{ij,k}-\Omega_{il}\Delta^l_{\;jk} +\Omega_{jl}\Delta^l_{\;ik}
(-1)^{\epsilon_i\epsilon_j}=0
\end{equation}
where $\Delta^i_{\;jk}$ ($\epsilon(\Delta^i_{\;jk})=
\epsilon_i+\epsilon_j+\epsilon_k)$ is a symmetric connection and the
symmetry property of $\Omega_{ij}$ (\ref{Omegainvsym}) was used. 

Let us introduce the curvature tensor of an antisymplectic
connection,
\begin{eqnarray}
\label{Rs}
{\cal R}_{ijkl}=\Omega_{in}{\cal R}^n_{\;jkl},\quad
\epsilon({\cal R}_{ijkl})=\epsilon_i+
\epsilon_j+\epsilon_k+\epsilon_l+1,
\end{eqnarray}
where ${\cal R}^n_{\;\;jkl}$ is given by (\ref{R}) with natural replacement
 $\Gamma^i_{\;jk}$ for $\Delta^i_{\;jk}$.
This leads to the following representation,
\begin{eqnarray}\label{r1}\nonumber
{\cal R}_{nljk}&=&-\Delta_{nlj,k}+\Delta_{nlk,j}(-1)^{\epsilon_j\epsilon_k}
+\Delta_{ink}\Delta^i_{\;lj}(-1)^
{\epsilon_n\epsilon_i+\epsilon_k(\epsilon_i+\epsilon_l+\epsilon_j)}\\
&&-\Delta_{inj}\Delta^i_{lk}(-1)^{\epsilon_n\epsilon_i+\epsilon_j
(\epsilon_i+\epsilon_l)}
\end{eqnarray}
where
\begin{eqnarray}\label{D1}
\Delta_{ijk}
=\Omega_{in}\Delta^n_{\;jk}\;,\quad\epsilon(\Delta_{ijk})
=\epsilon_i+\epsilon_j+\epsilon_k +1.
\end{eqnarray}

Using Eq. (\ref{D1}), the relation (\ref{Omegainvcon}) reads
\begin{eqnarray}
\label{Omegainvcon1} \Omega_{ij,k}=\Delta_{ijk}-
\Delta_{jik}(-1)^{\epsilon_i\epsilon_j}.
\end{eqnarray}
Furthermore, from Eq.~(\ref{r1}) it is follows that
\begin{eqnarray}
\label{Rans1} {\cal R}_{ijkl}=-(-1)^{\epsilon_k\epsilon_l}{\cal R}_{ijlk},
\end{eqnarray}
and, using (\ref{Rs}) and (\ref{Rjac}), one deduces the
Jacobi identity for ${\cal R}_{ijkl}$,
\begin{eqnarray}
\label{Rjac1} (-1)^{\epsilon_j\epsilon_l}{\cal R}_{ijkl}
+(-1)^{\epsilon_l\epsilon_k}{\cal R}_{iljk}
+(-1)^{\epsilon_k\epsilon_j}{\cal R}_{iklj}=0\,.
\end{eqnarray}
In addition, the curvature tensor ${\cal R}_{ijkl}$ is generalized
symmetric w.r.t. the first two indices,
\begin{eqnarray}
\label{Ras}
{\cal R}_{ijkl}=(-1)^{\epsilon_i\epsilon_j}{\cal R}_{jikl}.
\end{eqnarray}
In order to prove this, let us consider
\begin{eqnarray}
\label{com1}
\Omega_{ij,kl}=\Delta_{ijk,l}-
\Delta_{jik,l}(-1)^{\epsilon_i\epsilon_j}.
\end{eqnarray}
Then, using the relations
\begin{eqnarray}
\label{G1} \Delta_{ijk,l}=\Omega_{in}\Delta^n_{\;jk,l}
+\Omega_{in,l}\Delta^n_{\;jk}
(-1)^{(\epsilon_n+\epsilon_j+\epsilon_k)\epsilon_l}
\end{eqnarray}
and the definitions (\ref{r1}) and (\ref{com1}), we get
\begin{eqnarray}
\label{com1}
\nonumber
0&=&\Omega_{ij,kl}-(-1)^{\epsilon_k\epsilon_l}\Omega_{ij,lk}\\
\nonumber
&=&\Delta_{ijk,l}-
\Delta_{jik,l}(-1)^{\epsilon_i\epsilon_j}
-\Delta_{ijl,k}(-1)^{\epsilon_k\epsilon_l}+
\Delta_{jil,k}(-1)^{\epsilon_i\epsilon_j+\epsilon_k\epsilon_l}\\
&=&-{\cal R}_{ijkl}+(-1)^{\epsilon_i\epsilon_j}{\cal R}_{jikl}.
\end{eqnarray}
Moreover the curvature tensor obeys the identities 
\begin{eqnarray}
\label{Rjac3}
{\cal R}_{ijkl}
+(-1)^{\epsilon_l(\epsilon_i+\epsilon_k+\epsilon_j)}{\cal R}_{lijk}
+(-1)^{(\epsilon_k+\epsilon_l)(\epsilon_i+\epsilon_j)}
{\cal R}_{klij}+
(-1)^{\epsilon_i(\epsilon_j+\epsilon_l+\epsilon_k)}{\cal R}_{jkli}=0.
\end{eqnarray}
which have the same form as in the even case (\ref{Rjac2}).

Ricchi tensor can be defined by contracting two indices of curvature tensor
\begin{eqnarray}
\label{Ricdef}
{\cal R} _{ij}= {\cal R}^k_{\;ikj}\;(-1)^{\epsilon_k(\epsilon_i+1)}=
\Omega^{kn}{\cal R}_{nikj} (-1)^{\epsilon_i\epsilon_k+\epsilon_k+\epsilon_n},
\quad
\epsilon({\cal R}_{ij})=\epsilon_i+\epsilon_j\;.
\end{eqnarray}
Notice that in contrast with the even case Ricchi tensor (\ref{Richdef}) has no 
 a special symmetry properties.
The further contraction between antisymplectic tensor and Ricci tensor gives scalar curvature
\begin{eqnarray}
\label{Scalcur}
{\cal R} = \Omega^{ji}{\cal R}_{ij}\;(-1)^{\epsilon_i+\epsilon_j},\quad
\epsilon({\cal R})=1
\end{eqnarray}
which, in general, is not equal to zero. Notice that the scalar curvature
tensor squared is identically equal to zero, ${\cal R}^2=0$. 

If we identify $\Omega_{ij}$ with $\omega_{ij}$ then we can find coincidence antisymplectic supermanifolds and odd symplectic supermanifolds.

\section{Conclusion}

We have considered  possible generalizations of symplectic geometry on
manifolds to the supersymmetric case. In the even case there are
two scalar structures (an even closed non-degenerate differential
2-form and Poisson bracket) which can be used for clothing of a supermanifold.
When the Poisson bracket is non-degenerate and is constructed with the help of
tensor field inverse to a given symplectic structure, then
an even symplectic supermanifold and the non-degenerate Poisson
supermanifold coinñide. If, in addition, an even symplectic
supermanifolds is endowed with a symmetric connection (covariant
derivative) compatible with a given symplectic structure, one has an
even Fedosov supermanifolds which can be considered as
generalization of Fedosov manifolds \cite{fm}. In particular,
the scalar curvature tensor for even Fedosov supermanifolds vanishes.

In the odd case we have again two scalar structures (an odd closed
non-degenerate differential 2-form and the antibracket) for clothing
of a supermanifold. These structures lead to an odd symplectic
supermanifold and an antisymplectic supermanifold having the similar 
geometry.  The same statement is true if one equips these supermanifolds with  a symmetric connection
compatible with  given  structures.  The more important deference in contrast with the even case is non-triviality of the
scalar curvature tensor. Note that quite recently \cite{BB} the
scalar curvature tensor non-triviality was used in generalization of
the Batalin-Vilkovisky quantization scheme.

\section*{Acknowledgements}
The authors are grateful to K. Bering and I.V. Tyutin for
discussions. The work was partially supported by grant for LRSS,
project No.\ 4489.2006.2. The work of PML was also supported by the
INTAS grant, project INTAS-03-51-6346, the RFBR grant, project No.\
06-02-16346, the DFG grant, project No.\ 436 RUS 113/669/0-3 and
joint RFBR-DFG grant, project No.\ 06-02-04012.

\end{document}